\documentclass[useAMS,usenatbib]{mn2e}
\usepackage{amssymb,amsmath,epsfig,times,natbib,multirow}

\long\def\symbolfootnote[#1]#2{\begingroup%
\def\thefootnote{\fnsymbol{footnote}}\footnote[#1]{#2}\endgroup} 

\newcommand{\nustar}{\emph{NuSTAR}}
\newcommand{\swift}{{\it Swift}}
\newcommand{\suzaku}{{\it Suzaku}}

\title[Extreme Relativistic Effects in Mrk 335]{The \nustar{} spectrum of Mrk 335: Extreme relativistic effects within 2 gravitational radii of the event horizon?}

\author[M. L. Parker et al.]{M. L. Parker,$^1$\thanks{Email: mlparker@ast.cam.ac.uk}
  D. R. Wilkins,$^{1,2}$
  A. C. Fabian,$^{1}$
  D. Grupe,$^{3}$
  T. Dauser,$^{4}$
  G. Matt,$^{5}$ \newauthor
  F. A. Harrison,$^{6}$
  L. Brenneman,$^{7}$
  S. E. Boggs,$^{8}$                                                                               
  F. E. Christensen,$^{9}$                                                                         
  W. W. Craig,$^{10,11}$ \newauthor
  L. C. Gallo,$^{2}$                                                                               
  C. J. Hailey,$^{11}$                                                                             
  E. Kara,$^{1}$                                                                                   
  S. Komossa,$^{12}$                                                                               
  A. Marinucci,$^{5}$                                                                              
  J. M. Miller,$^{13}$ \newauthor
  G. Risaliti,$^{7,14}$                                                                            
  D. Stern,$^{15}$                                                                                 
  D. J. Walton$^{6}$ and
  W. W. Zhang$^{16}$\\                                   
  $^{1}$Institute of Astronomy, Madingley Road, Cambridge, CB3 0HA\\                               
  $^{2}$Saint Mary's University, 923 Robie Street, Halifax, NS B3H 3C3, Canada\\                   
  $^{3}$Department of Astronomy and Astrophysics, Pennsylvania State University, 525 Davey Lab, University Park, PA 16802, USA\\                                                                      
  $^{4}$Dr Karl Remeis-Observatory and Erlangen Centre for Astroparticle Physics, Sternwartstr. 7, D-96049 Bamberg, Germany\\                                                                         
  $^{5}$Dipartimento di Matematica e Fisica, Universit\`{a} degli Studi Roma Tre, via della Vasca Navale 84, 00146 Roma, Italy\\                                                                      
  $^{6}$California Institute of Technology, 1200 East California Boulevard, Pasadena, CA 91125, USA\\                                                                                                 
  $^{7}$Harvard-Smithsonian Center for Astrophysics, 60 Garden St, Cambridge, MA 02138, USA\\      
  $^{8}$Space Sciences Laboratory, University of California, Berkeley, 7 Gauss Way, Berkeley, CA 94720-7450, USA\\                                                                                    
  $^{9}$Danish Technical University, DK-2800 Lyngby, Denmark\\                                     
  $^{10}$Lawrence Livermore National Laboratory, Livermore, CA, USA\\                              
  $^{11}$Columbia University, New York, NY 10027, USA\\                                            
  $^{12}$Max-Planck-Institut f\"ur Radioastronomie, Auf dem H\"ugel 69, D-53121 Bonn, Germany\\    
  $^{13}$Department of Astronomy, University of Michigan, 500 Church Street, Ann Arbor, MI 48109-1042, USA\\                                                                                          
  $^{14}$INAF--Osservatorio Astrofisico di Arcetri, Largo Enrico Fermi 5, 50125 Firenze, Italy\\
  $^{15}$Jet Propulsion Laboratory, California Institute of Technology, 4800 Oak Grove Drive, Pasadena, CA 91109, USA\\
  $^{16}$NASA Goddard Space Flight Center, Greenbelt, MD 20771, USA\\
 }
\date{}

\begin{document}

\maketitle

\begin{abstract}
We present 3--50~keV \nustar{} observations of the AGN Mrk 335 in a very low flux state. The spectrum is dominated by very strong features at the energies of the iron line at 5--7~keV and Compton hump from 10--30~keV. The source is variable during the observation, with the variability concentrated at low energies, which suggesting either a relativistic reflection or a variable absorption scenario.
In this work we focus on the reflection interpretation, making use of new relativistic reflection models that self consistently calculate the reflection fraction, relativistic blurring and angle-dependent reflection spectrum for different coronal heights to model the spectra. 
We find that the spectra can be well fit with relativistic reflection, and that the lowest flux state spectrum is described by reflection alone, suggesting the effects of extreme light-bending occurring within $\sim2$ gravitational radii ($R_\textrm{G}$) of the event horizon.
The reflection fraction decreases sharply with increasing flux, consistent with a point source moving up to above 10 $R_\textrm{G}$ as the source brightens. We constrain the spin parameter to greater than 0.9 at the 3$\sigma$ confidence level.
By adding a spin-dependent upper limit on the reflection fraction to our models, we demonstrate that this can be a powerful way of constraining the spin parameter, particularly in reflection dominated states.
We also calculate a detailed emissivity profile for the iron line, and find that it closely matches theoretical predictions for a compact source within a few $R_\textrm{G}$ of the black hole.
\end{abstract}

\begin{keywords}
galaxies: active -- galaxies: Seyfert -- galaxies: accretion -- galaxies: individual: Mrk 335
\end{keywords}

\section{Introduction}

Several active galactic nuclei (AGN) exhibit large flux variations, in which the source spectrum usually becomes much harder in very low flux states. This behaviour can be caused by two different mechanisms: material passing across the line of sight to the AGN absorbing the light from the source \citep[e.g. NGC~1365,][]{Risaliti13}, or the source switching to a reflection dominated state \citep[e.g. 1H~0707-495][]{Fabian12}. Distinguishing between these two cases can be difficult, particularly when data is restricted to $E \lesssim$ 10~keV.

The \emph{Nuclear Spectroscopic Telescope Array} \citep[\nustar{},][]{Harrison13} is the first focusing high energy (3--79 keV) X-ray telescope in orbit. Because of this, it is uniquely suited to studying reflection spectra in AGN and X-ray binaries (XRBs), as it can simultaneously observe the broad iron line from $\sim$3--7~keV and the Compton hump, prominent from $\sim$10--30~keV, \citep[e.g.][]{Risaliti13,Miller13,Tomsick13,Marinucci14a,Marinucci14b} with good enough spectral resolution to constrain the iron line profile and hence the spin and emissivity profile. This constraint on the shape of the high energy spectrum can enable the effects of absorption and reflection to be disentangled. Recent work on NGC~1365 with \emph{NuSTAR} \citep{Walton14} has demonstrated that when intervening cold absorption is taken into account the fluxes of the iron line and Compton hump are correlated. These features show no systematic correlation with the absorption in the source, and are found in all flux states, pointing to an origin unrelated to absorption.

Mrk~335 is a bright narrow-line Seyfert 1 (NLS1) galaxy at redshift $z=0.025785$, with $\log(M_\textrm{BH})=7.15\pm0.12$ \citep{Peterson04} and evidence of ionised absorption in the X-ray and UV \citep{Longinotti13}. Mrk~335 has been observed to go into extremely low flux states \citep{Grupe07} where the soft X-ray flux drops by a factor of up to $\sim 30$, while the hard flux only drops by a factor of $\sim$~2. \citet{Grupe08}, using \emph{XMM-Newton}, found that the spectrum of Mrk 335 in its lowest state can be described equally well by either the effects of absorption or by a high reflection fraction, with no clear way to distinguish between the two models. \citet{Gallo13} analysed the spectrum of Mrk 335 in an intermediate flux state from 2009 \citep{Grupe12}, and found that the complex variability can be well described using a blurred reflection model without invoking variable absorption. Extremely low flux states have also been observed in several other AGN, such as NGC~4051 \citep{Guainazzi98,Uttley99,Lobban11}, Fairall~9 \citep{Lohfink12}, PG~2112+059 \citep{Schartel10}, PG~0844+349 \citep{Gallo11}, PHL~1092 \citep{Miniutti12} and Mrk~1048 (Parker et al., submitted).

X-ray reflection is observed when emission from a hot X-ray emitting corona is reflected off an optically thick accretion disk, producing both fluorescent lines and a backscattered continuum. These features can then be smeared by general relativistic effects close to the event horizon \citep{Fabian89,Laor91}. This phenomenon is well documented in AGN and XRBs \citep[e.g.][]{Fabian13}. Relativistic reflection spectra are characterised by a soft excess of blurred emission lines at low energies, a broad iron K line at $\sim6.4$ keV, and a Compton hump at high energies ($\gtrsim 10$ keV, peaking around 20--30 keV).
By investigating the properties of the relativistic blurring, we can investigate the geometry of the corona. The corona has been found to be compact using several different methods \citep[e.g. time lags and microlensing,][respectively]{Zoghbi10,Dai10}, and the height of the corona above the disk can be constrained by looking at the iron line profile. The closer the source of the coronal emission is to the disk, the more the reflected emission is dominated by the inner disk, and the more extreme the blurring \citep{Wilkins12,Dauser13,Fabian14}.
A characteristic of relativistic reflection from a compact corona close to the event horizon is that changes in the position of the coronal X-ray source can result in large changes in the continuum flux, with only a small difference in the reflected spectrum. This is a consequence of light bending, used by \citet{Miniutti03} to explain the lack of variability in the reflection spectrum of MCG--06-30-15, later expanded by \citet{Miniutti04}. In this model the reflection fraction is set by the height of the illuminating source above the disk. When the source is high above the disk, approximately half of its emission falls onto the disk and half escapes to infinity. However, when the corona is within a few $R_\textrm{G}$ of the disk, general relativistic light bending results in a smaller fraction of the emitted photons escaping to infinity, while a roughly constant fraction falls onto the disk and a larger number fall into the event horizon \citep[Dauser et al., submitted]{Fukumura07}. 

Arguably the strongest evidence for the relativistic reflection interpretation of the spectrum of Mrk~335 comes from reverberation lags. \citet{Kara13} found a high frequency lag of $\sim150$~s between the continuum dominated energy bands and those which should include a large relativistic reflection component (the iron line and soft excess). This lag corresponds to the light travel time from the corona to the accretion disk, which delays the reflected spectrum. In the absorption model, lags arise as a consequence of scattering from distant material \citep[e.g.][]{Legg12}, and as such there is no explanation for the broad iron line profile present in the lag energy spectrum of this and other similar sources. Additionally, \citet{Walton13} showed that the low frequency lag commonly seen in AGN cannot arise from distant reprocessing, and must instead be intrinsic to the source.
Fe K lags appear to increase in amplitude with BH mass, following the same trend as found for lags of the soft excess \citep{DeMarco13}. The 150~s amplitude of the lag in Mrk~335 corresponds to the light travel time of $\sim2$~$R_\textrm{G}$, suggesting that the continuum source is located very close to the black hole and that relativistic effects are likely to be very important.

In this paper we present the first broad-band 0.5--50~keV observation of Mrk~335 in the low state, which reveals a spectrum consistent with being almost entirely dominated by reflected emission from the inner accretion disk.

\section{Observations and Data Reduction}
In May 2013, \emph{Swift} monitoring found Mrk 335 in a very low flux state. A 300 ks \emph{Suzaku} Target of Opportunity observation was triggered, and we also performed follow up with \nustar{}. The \emph{Suzaku} observation took place in June, and partially overlaps with those of \nustar{}. The \emph{Suzaku} data will be presented in a separate paper (Gallo et al., in preparation), focusing on a detailed comparison of the reflection and absorption models. 

\subsection{NuSTAR}
\label{section_nustarreduction}
Mrk 335 was observed twice with \nustar{}, in June and July 2013, for $\sim80$ and $\sim200$~ks of elapsed time (the net exposure time is approximately 50 per cent of the total observation length, due to the low Earth orbit). The first observation was split into two separate observation IDs, with approximately the same exposure time in each. The three observation IDs used are shown in Table~\ref{obstable}. 

The data were reduced and filtered for background flares using the \nustar\ Data Analysis Software (NUSTARDAS) 1.3.1. \nustar caldb version 20131223 was used for all instrumental responses.  All spectra and lightcurves are extracted from the cleaned event files using \textsc{nuproducts}. We used $65^{\prime\prime}$ extraction regions for all source and background spectra, and the spectra were binned to have a signal to noise ratio of five after background subtraction and to oversample the spectra by a factor of 3. The background regions were selected from the same detector, in a region free of point sources. The resulting spectra are source dominated until $>40$~keV, and are consistent between the two focal plane modules (FPMA and FPMB).

\begin{table}
\begin{tabular}{l l p{1.4cm} l}
\hline
Obs. ID & Start date & On-source time (ks)& FPMA count rate (s$^{-1}$)\\
\hline
\hline
60001041002 & 2013/06/13 & 21.3 & $0.095\pm0.02$\\
60001041003 & 2013/06/13 & 21.5 & $0.127\pm0.03$\\
60001041005 & 2013/06/25 & 93.0 & $0.180\pm0.02$\\
\hline
\end{tabular}
\caption{\nustar{} observations of Mrk 335 used in this analysis. The total count rate for the whole \nustar{} energy band is given for the FPMA detector, which has a slightly higher count rate than FPMB.}
\label{obstable}
\end{table}

\begin{figure}
\includegraphics[width=\linewidth]{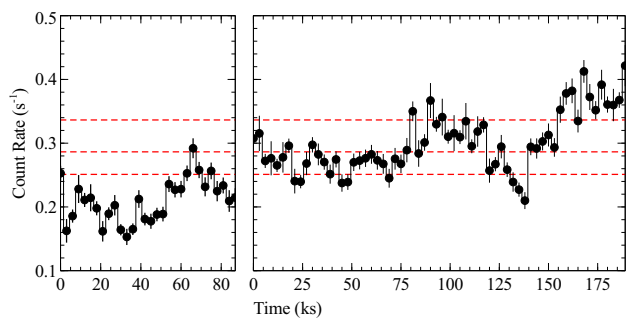}
\caption{Background subtracted light curves of the two \nustar{} observations of Mrk 335. The different flux levels used to extract spectra are shown by the horizontal lines. Note that for technical reasons the first observation was split into two separate observation IDs. For this figure, we sum the FPMA and FPMB count rates, and use 3~ks bins.}
\label{lightcurve}
\end{figure}

To investigate flux-dependent variability in Mrk~335, we split the data into four separate flux states: very low, low, high and very high. The spectra are selected by filtering the 1~ks binned FPMA light curve such that the total number of counts is approximately the same (25 per cent of the total counts) in each spectrum, as shown in Fig.~\ref{lightcurve}. These spectra are then binned as discussed above.

\begin{figure}
\includegraphics[width=\linewidth]{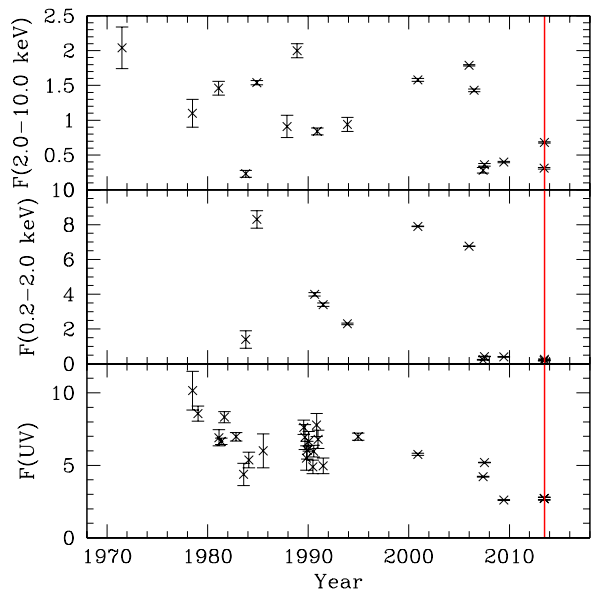}
\caption{Long term lightcurve of Mrk 335 as in \citet{Grupe08}. Top: 2--10~keV light curve, starting from the \emph{Uhuru} detection in 1971, and ending with the two \nustar{} observations, with the 2--3~keV flux extrapolated from the best fitting 3--50~keV model. Middle: the 0.2--2~keV lightcurve, starting from the 1983 \emph{EXOSAT} observation and ending the \emph{Swift} observation from the same day as the first \nustar{} observation. Bottom: UV lightcurve, measured with \emph{IUE}, \emph{HST}, \emph{XMM-Newton} OM and \emph{Swift} UVOT. The UV data are not corrected for Galactic reddening. The time of the \nustar{} observations is marked by the red vertical line. All fluxes are given in units of $10^{-11}$~erg~s$^{-1}$~cm$^{-2}$.}
\label{longterm_lightcurve}
\end{figure}

Fig.~\ref{longterm_lightcurve} shows the long term lightcurve of Mrk~335, from its detection in 1971 to the latest \nustar{} observation. The observations used in this plot are described in Table 3 of \citet{Grupe08}, with additional recent data from \emph{Swift}, \emph{XMM-Newton} and \nustar{} added.

\begin{table}
\centering
\begin{tabular}{c c c}
\hline
Spectrum & FPMA count rate (s$^{-1}$) & Exposure time (ks)\\
\hline
\hline
1 & 0.095$\pm$0.001 & 46.7\\
2 & 0.130$\pm$0.002 & 36.9\\
3 & 0.150$\pm$0.002 & 30.2\\
4 & 0.181$\pm$0.003 & 22.0\\
\hline
\end{tabular}
\caption{Count rates and exposure times for the four \nustar\ spectra, covering different flux states. Spectra are selected so as to have approximately the same total counts.}
\label{spectable}
\end{table}

\subsection{Swift}
Mrk 335 has been a target of a regular monitoring campaign by \swift\ \citep{Grupe12}
after it was discovered in an unusual low X-ray flux state \citep{Grupe07}. 
While our usual monitoring program had a four day cadence,
during the \nustar\ observations we increased the frequency to daily observations.
In order to obtain a better signal to noise ratio in the two spectra during the
\nustar\ observations we merged the data of 2013 June 11-14 and June 25-27, which are consistent in flux and photon index. We then merged these two spectra to give a single average spectrum, which should be approximately consistent with the average \nustar\ spectrum. We use a simple binning of 20 counts per bin, and fit the XRT data from 0.3--7~keV.

The observations with the \swift\ X-ray Telescope 
\citep[XRT,][]{Burrows05} were all performed in photon counting mode 
\citep{Hill04}.
XRT data were reduced with the the task {\it xrtpipeline} version 0.12.6., 
which is included in the HEASOFT package 6.12. For each day we extracted a
spectrum using XSELECT. Source counts were extracted in a circular region with a
radius of 94$^{''}$ and the background counts from a source-free
region with a radius 295$^{''}$. We created auxiliary response files (ARF) for
each of these spectra using the FTOOL {\it xrtmkarf}. These ARFs were the
combined into a single ARF per observation period using the FTOOL {\it addarf}.
We used the most recent response file  {\it swxpc0to12s6\_20010101v013.rmf}.

All spectral analysis was performed using \textsc{xspec} version 12.8.1, and all errors are quoted at 1$\sigma$ confidence levels unless otherwise specified.

\section{Spectral Modelling}

Fig.~\ref{ratio} shows the ratio of the 100 ks observation by \nustar\ to a power law, fit from 3--4, 8--10, and 40--50~keV to exclude bands where strong reflection or absorption features may be present. The photon index of the power law is $1.72\pm0.03$. This model is clearly a poor fit, and separate narrow (at $\sim6$keV) and broad (4--7 ~keV) features can be seen, along with a high energy excess peaking at $\sim20$~keV. We note that the residuals in Fig.~\ref{ratio} do not represent the true shape of the component causing the excess: for that the underlying continuum needs to be determined, which is likely to be steeper than that found from this fit. In this and all following fits to the \nustar\ data alone, we neglect Galactic absorption as its effect is negligible in the \nustar\ band for this source. To thoroughly investigate the spectrum of this source, we first consider the time-averaged spectrum, then simultaneously fit the spectra from the four different flux-states.

In all our fits to the spectra of Mrk~335, we include a single highly ionised warm absorption zone. This absorption is to allow for the potential narrow iron absorption features in the 7--8~keV band, which are also seen in NGC~1365 \citep{Risaliti13}. In all cases, this absorption zone is too highly ionized to have any significant effect on the continuum or broad iron line fits. While \citet{Longinotti13} found evidence for three warm absorption zones, the effect of the two lower ionisation zones is negligible over the \nustar\ band. \footnote{We note that the predictions for the iron absorption features at $\sim$7--8~keV differs between the \textsc{xstar} grids used here and the \textsc{xabs} SPEX model used in \citet{Longinotti13}. At the ionisation states reported by \citeauthor{Longinotti13}, \textsc{xabs} does not predict any features over this band, while \textsc{xstar} does. It is therefore unclear whether this represents the same absorption zone or another, higher ionisation zone.}

\subsection{Time-Averaged Spectrum}
\label{section_avg}

\begin{figure}
\includegraphics[width=\linewidth]{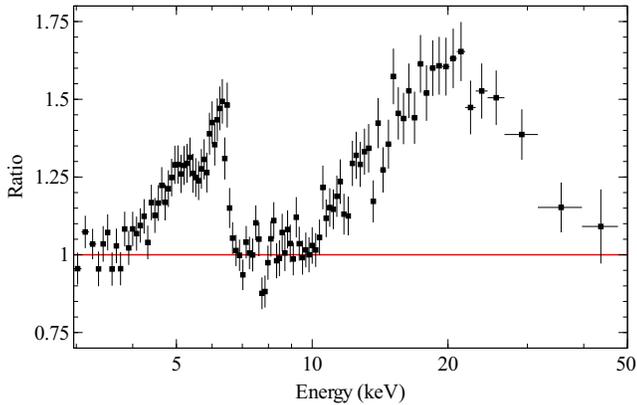}
\caption{Ratio of the 100 ks \nustar{} observation of Mrk 335 to a power law, fit from 3--4, 7--10 and 40--50~keV. A narrow iron line is visible at just above 6~keV, as well as a broader feature extending from around 4--7~keV. Above 10~keV, a large excess is visible. Possible small absorption features are visible at $\sim$7 and 8 keV. The FPMA and FPMB spectra are grouped in this plot for clarity, but are kept separate for spectral fitting. Data are re-binned in \textsc{xspec} for visual clarity.}
\label{ratio}
\end{figure}

\begin{figure}
\includegraphics[width=\linewidth]{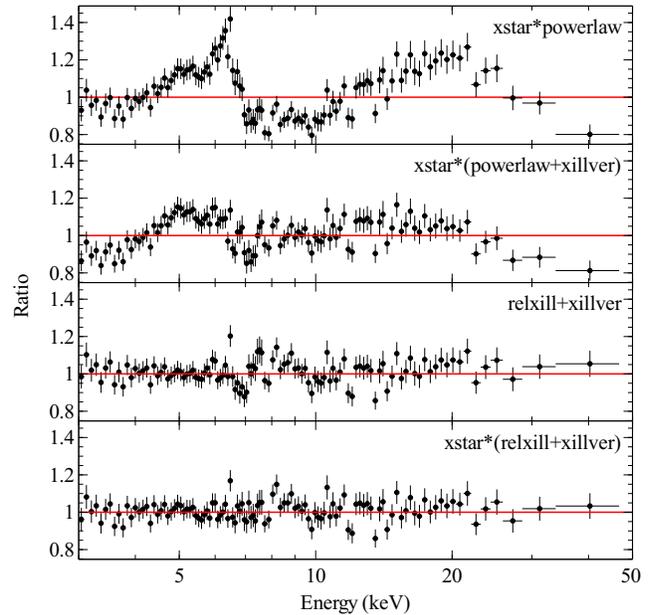}
\caption{Data/model ratios for the models (1,2 and 4 from Table~\ref{avgfittable}) fit to the time-averaged spectrum. We also show (3rd row) the residuals from fitting model 4 without a warm absorber. Data are rebinned and grouped in \textsc{Xspec} for clarity, but both detectors are fit independently and with the binning described in the text. All models are fit from 3--50~keV, where the spectrum is source-dominated.}
\label{avgspecrats}
\end{figure}

We initially investigate a simple fit using a power law plus distant reflection alone, as would be expected if the intrinsic source flux had dropped sharply so that distant material contributes significantly to the observed spectrum. We model the distant reflection using the \textsc{xillver} model \citep{Garcia13}, fixing the incident photon index to the same value as the power law but leaving all other parameters free. The resulting fit is poor ($\chi^2_\nu=584/408=1.43$) (residuals for this and other fits to the average spectrum are shown in Fig.~\ref{avgspecrats}). While this removes the narrow line feature shown in Fig.~\ref{ratio} and flattens out the high energy feature, significant residuals remain from 4--6~keV and above 20~keV. The parameters for this model and all others fit to the average spectrum are given in Table~\ref{avgfittable}.

To examine the blurred reflection spectrum of Mrk~335, we use three different relativistic reflection models: the conventional \textsc{reflionx} convolved with \textsc{relconv} \citep{Ross05,Dauser10}, and the new \textsc{relxill} and \textsc{relxilllp} models \citep{Dauser13,Garcia14}. The \textsc{relxill} models self consistently calculate the angle dependent reflection spectrum from the disk, and in the case of \textsc{relxilllp} the reflection fraction and line profile is calculated as a function of source height, assuming a simple lamppost geometry. By comparing the three models, we aim to establish the robustness of our parameter estimates, as well the plausibility of simple light bending models. We include an additional constraint to the two non-lamppost models, limiting the reflection fraction for each spin value to be less than the maximum possible reflection fraction for a lamppost geometry at that spin. This constraint rules out unphysical models, and can give much more precise measurements of spin in reflection-dominated spectra (for more detailed discussion of this method, see Dauser et al., submitted).

\begin{figure*}
\includegraphics[width=8cm]{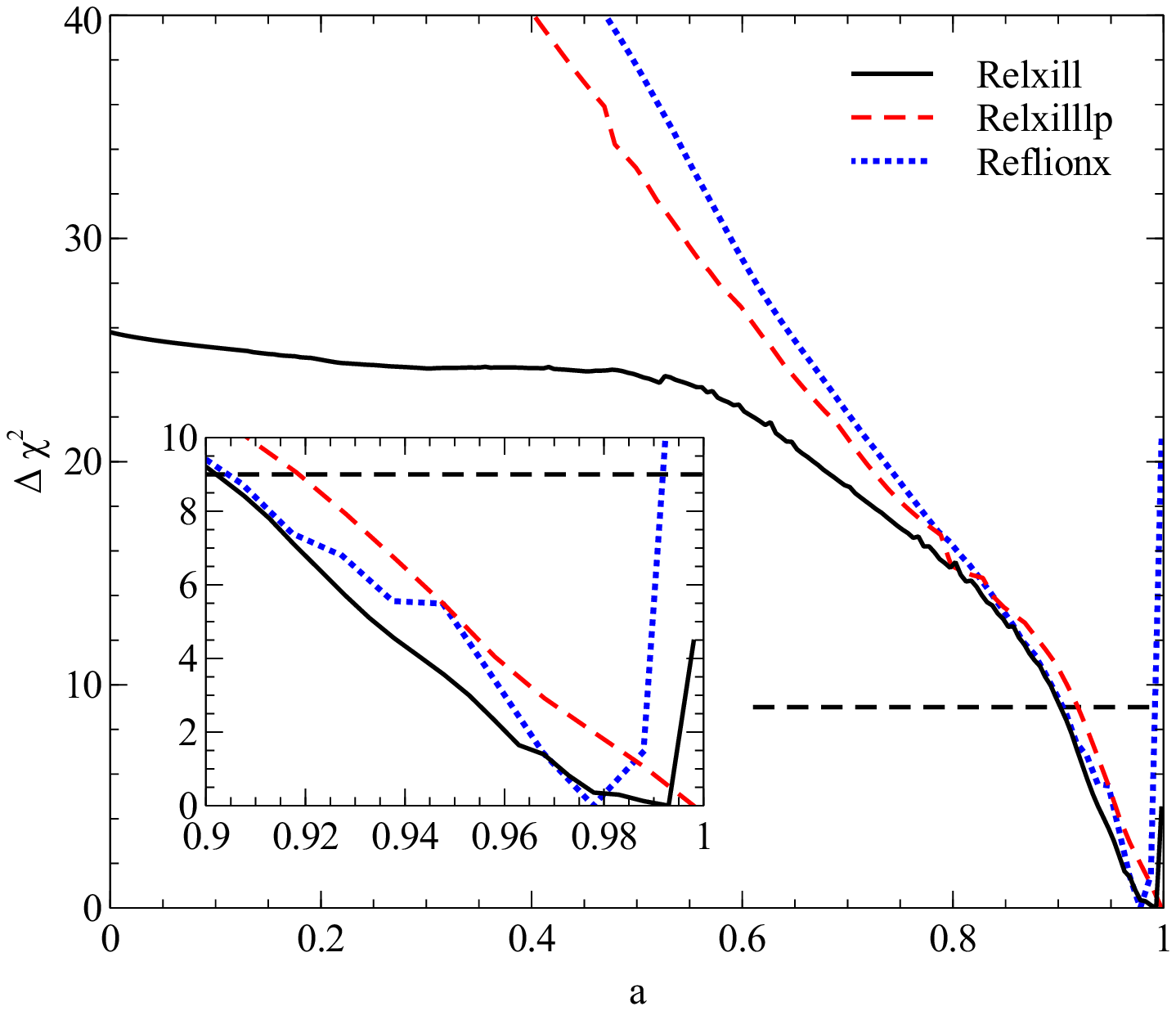}
\includegraphics[width=8cm]{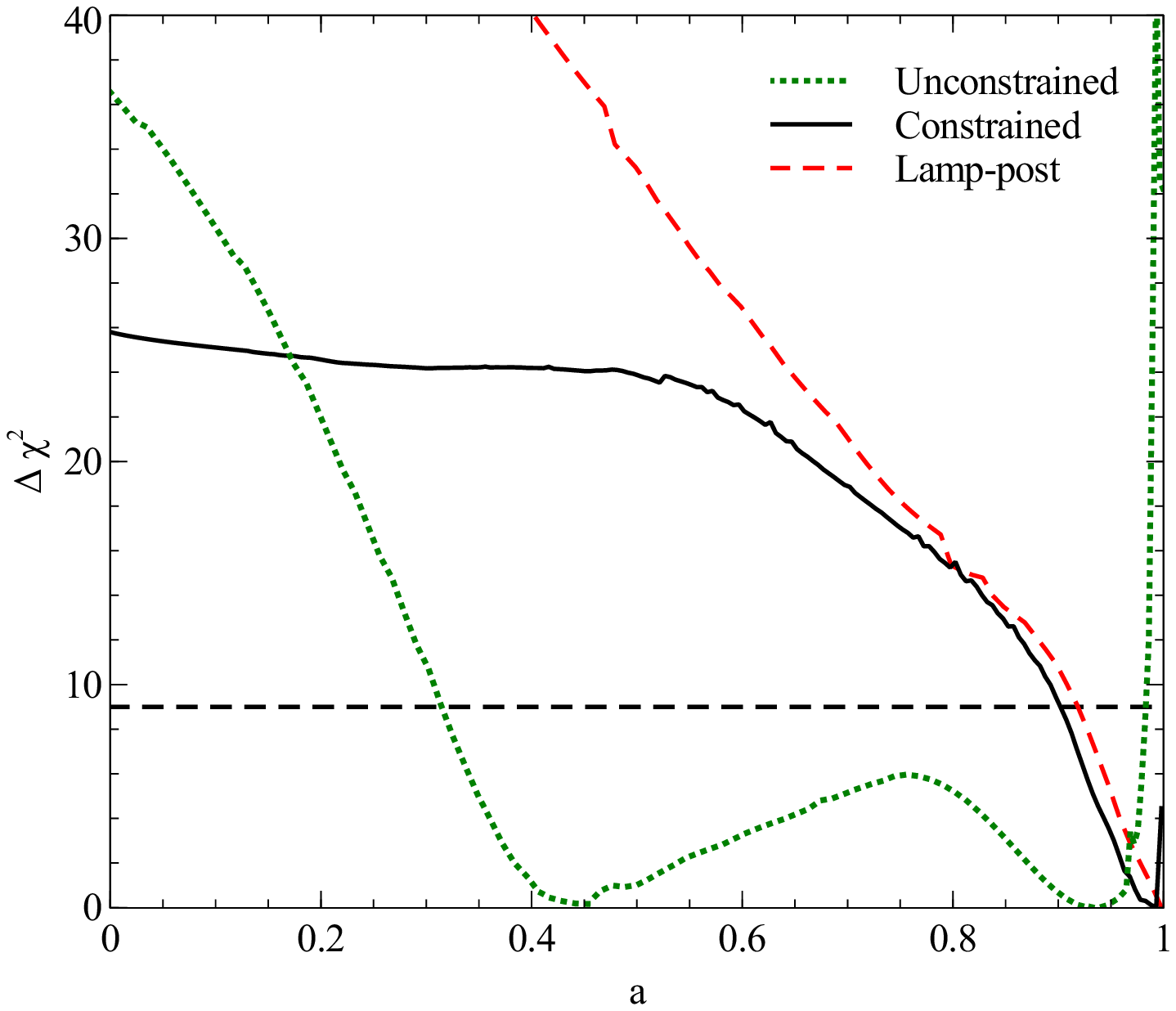}
\caption{Left: $\Delta\chi^2$ contours for the three relativistic reflection models fit to the time-averaged spectrum. The inset shows the $a=0.9$--1.0 region, and the dashed line in the main figure shows the 3$\sigma$ confidence limit for a single parameter of interest. The agreement between the models is very good, although the lamppost model hits the maximum spin limit without any upturn in $\chi^2$. Right: as left, but for the \textsc{Relxill} in three different cases: without any constraint on the reflection fraction, with the constraint added, and the full lamp-post solution. The constrained and lamp-post lines are the same as in the left panel.}
\label{avgchisquareds}
\end{figure*}

All three of the blurred reflection models give excellent fits (see Table~\ref{avgfittable}), and all three require high spin (Fig.~\ref{avgchisquareds}, left panel). The high reflection fractions are consistent between the models, as are the parameters of the warm absorber, however, there appear to be systematic differences between some of the other parameters. The ionisation of the reflector differs somewhat between the models, but this is expected as the differences caused by ionisation changes occur largely at lower energies, outside the \nustar\ band. The \textsc{reflionx} model gives a softer power law, steeper emissivity profile, and higher inclination than the \textsc{relxill} models, while \textsc{relxilllp} requires a higher iron abundance.

In the right panel of Fig.~\ref{avgchisquareds} we show the improvement in the spin constraint obtained by restricting the reflection fraction to physical values. Without this constraint, there are two acceptable values of the spin at the 2$\sigma$ level, at $\sim0.4$ and $\sim0.9$. However, the low spin solution still requires a high reflection fraction ($R>3$) and a steep emissivity profile ($q>5$). This is unphysical, as a steep emissivity profile and reflection fraction significantly greater than one require a large degree of light-bending, which can only be achieved if the inner disk extends very close to the black hole, which in turn requires high spin. By forcing the model to remain within regions of parameter space where the reflection fraction is plausible, we obtain a much tighter constraint, with all spins below $a=0.9$ ruled out at the three sigma level. This also gives a much better agreement among the models.

\begin{table*}
\begin{tabular}{c c c c c c c c c c c}
\hline
Model & $N_\textrm{H}$ & $\log(\xi_\textrm{abs})$ & $\Gamma$ & $A_\textrm{Fe}$ &$\theta$&$a$&$R/h$&$\log(\xi_\textrm{ref})$&$q_\textrm{in}$ &$\chi^2_\nu/\textrm{d.o.f.}$\\
& $10^{22}$~cm$^{-2}$ & log(erg cm s$^{-1}$)& & solar&degrees&&- /$R_\textrm{G}$&log(erg cm s$^{-1}$)&\\
\hline
\hline
1 & 10.5 & 4.33 & 1.61 & 5.0&-&-&-&-&-&2.28/409\\
2 & $33.5^{+1.3}_{-2.1}$ & $3.9\pm0.1$ & $2.32^{+0.03}_{0.05}$ & $<1.06$ & $<89$&-&-&-&-& 1.25/408\\
3 & $10.0^{+1.6}_{-4.8}$ & $>3.8$ & $2.32^{+0.03}_{-0.06}$&$1.99^{+0.33}_{-0.07}$ & $64\pm1$ &$0.98\pm0.01$&$>5.08$& $1.6\pm0.1$ &$>8.2$& 1.00/404\\
4 & $14.0^{+2.5}_{-7.8}$ & $>3.7$ & $1.95^{+0.03}_{-0.04}$ & $1.8\pm0.5$& $30_{-6}^{+3}$ &$0.99^{+0.01}_{-0.02}$& $9.2^{+0.6}_{-3.6}$& $2.78\pm0.03$ &$3.5\pm0.5$&1.00/404 \\
5 & $<17$ &$>3.4$ & $2.0^{+0.1}_{-0.2}$ & $3.0\pm0.5$& $31^{+8}_{-6}$ & $>0.97$ &$3.1\pm0.4$&$2.4^{+0.4}_{-0.2}$&-&1.02/404 \\
\hline
\end{tabular}
\caption{Model parameters for fits to the time-averaged \nustar\ spectrum. All models include absorption by a single, highly ionised absorption zone, modelled with \textsc{xstar}. Model 1 is a simple power law, with no reflection (errors are not shown for this model, as the reduced $\chi^2$ was too high for \textsc{xspec} to calculate parameter errors); 2 is the distant reflection (\textsc{powerlaw+xillver}) model and models 3, 4 and 5 are the \textsc{relconv*reflionx, relxill} and \textsc{relxilllp} models, respectively. Note that for model 5 we give the source height $h$, rather than the reflection fraction $R$, as this model sets $R$ based on $h$.}
\label{avgfittable}
\end{table*}

Based on these fits we conclude that blurred reflection models do an excellent job of modelling the time-averaged \nustar\ spectrum of the Mrk~335 low state, requiring a high-reflection fraction and correspondingly high spin.

\subsection{Broad-band model}

\begin{figure*}
\includegraphics[width=12cm]{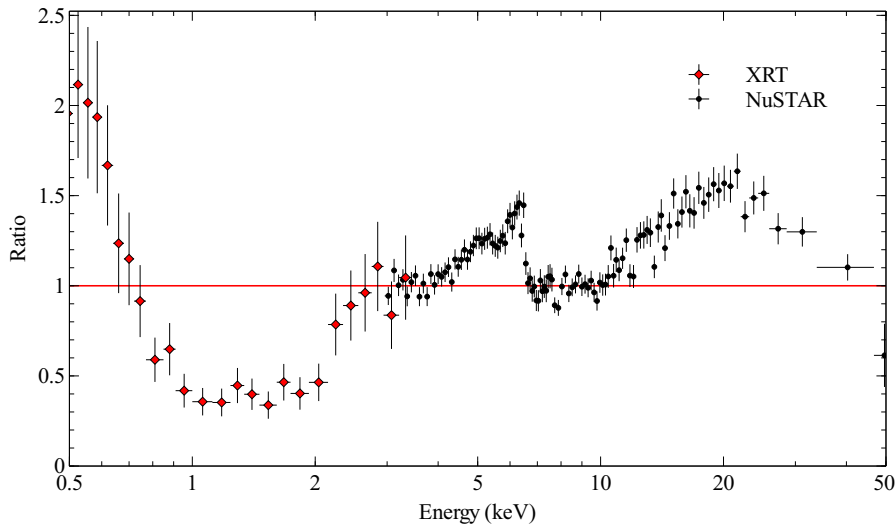}
\caption{\nustar\ and XRT residuals from the average spectra fit with a power law, modified by Galactic absorption, over the 3--4, 8--10 and 40--50~keV bands. The \nustar\ data are rebinned slightly in \textsc{xspec} and we ignore the XRT data above 3.5~keV for clarity (the XRT data are fit from 0.3--7~keV). As in Fig.~\ref{ratio} the broad iron line and Compton hump are visible, but the XRT data reveals the presence of warm absorption and a soft excess as well.}
\label{xrtrat}
\end{figure*}

Having established a best fit relativistic reflection model for the average \nustar\ spectrum, we then extend this to the soft band using the average \swift\ XRT spectrum. In Fig.~\ref{xrtrat} we show the residuals from a power law, modified by Galactic absorption with $N_\textrm{H}$ of $3.56\times 10^{20}$~cm$^{-2}$ \citep{Kalberla05}, fit to the 3-4,8-10 and 40--50~keV bands (where strong reflection and absorption features should not be present). In addition to the broad line and high energy hump, this figure shows a strong soft excess and clear warm absorption.

For the fit to this combined dataset the warm absorption becomes more important, so we include an additional, lower ionisation, \textsc{xstar} grid to model this, as well as Galactic absorption. We use the model \textsc{tbabs*xstar*xstar*(relxill+xillver)} rather than independently fitting all three reflection models as the \swift\ data quality is not sufficient to distinguish between them. 

\begin{figure}
\includegraphics[width=\linewidth]{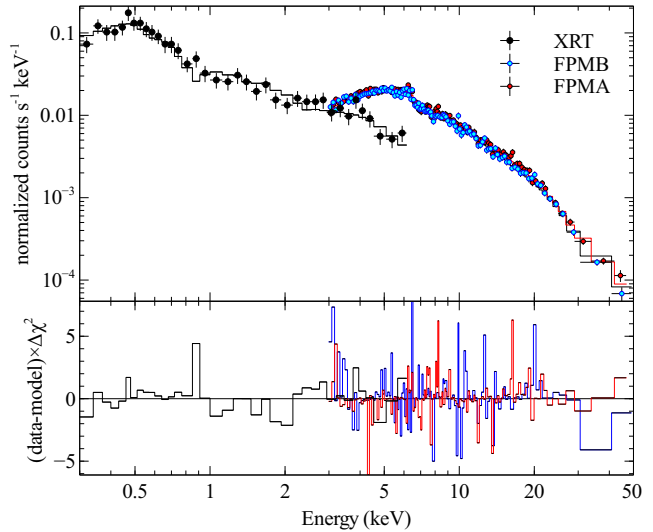}
\caption{Data and residuals from fitting a blurred reflection plus power law model, including warm absorption, to the combined XRT and \nustar\ spectra. Note that we plot $\delta\chi^2$ rather than a ratio, so that the residuals from both datasets can be seen on the same axes.}
\label{xrtfig}
\end{figure}

We find an excellent fit ($\chi^2_\nu=446/440=1.01$) to the broad-band spectrum with this model, despite not allowing for spectral variability between the \nustar\ and \swift\ observations or differences in the calibration. The model parameters are shown in Table~\ref{jointfittable}. We stress that we are approximating the multiple warm absorption zones which affect the soft band \citep{Longinotti13} with a single zone, as the XRT spectrum does not have enough signal or resolution to constrain multiple zones. This approximation does a good job of fitting the data, but the parameters returned should be treated with a degree of caution.

\begin{table}
\centering
\begin{tabular}{l c c}
\hline
Component & Parameter & Value \\
\hline
\hline
Absorber 1 & $N_\textrm{H}$ & $7.35\pm0.07$\\
& $\log(\xi)$ & $2.04\pm0.04$\\
Absorber 2 & $N_\textrm{H}$ & $10\pm4$\\
& $\log(\xi)$ & $>3.9$\\
Powerlaw & $\Gamma$ & $1.85\pm0.04$\\
Reflection & $q_\textrm{in}$ & $4.72_{-0.36}^{+0.25}$\\
& $a$ & $0.99^{+0.01}_{-0.03}$\\
& $A_\textrm{Fe}$ & $3.0_{-0.8}^{+0.6}$ \\
&$\log(\xi_\textrm{ref})$&$3.02\pm0.02$\\
&$R$&$5_{-2}^{+3}$\\
&$\theta$&$25\pm5$\\
\hline
\end{tabular}
\caption{Parameters estimates from the fit to the combined XRT and \nustar\ spectra, fit with a power law plus reflection model, and two warm absorbing zones (\textsc{xstar*xstar*[powerlaw+relxill+xillver]}). All units are the same as in Table~\ref{avgfittable}.}
\label{jointfittable}
\end{table}

The parameters returned from the joint fit are in general compatible with those found from the \nustar\ data alone, and a high reflection fraction is still required to fit the data. We conclude that the broad-band average spectrum is well modelled using a reflection-dominated solution.

\subsection{Flux-Resolved Spectra}
We now focus on the flux-resolved analysis. For this, we focus on the \nustar\ data alone, as the XRT data are not of high enough quality and do not cover the same time intervals as the \nustar\ observations. As discussed in \S~\ref{section_nustarreduction} we use four spectra, selected to have the same total number of counts, from four different flux levels.

\begin{figure}
\includegraphics[width=\linewidth]{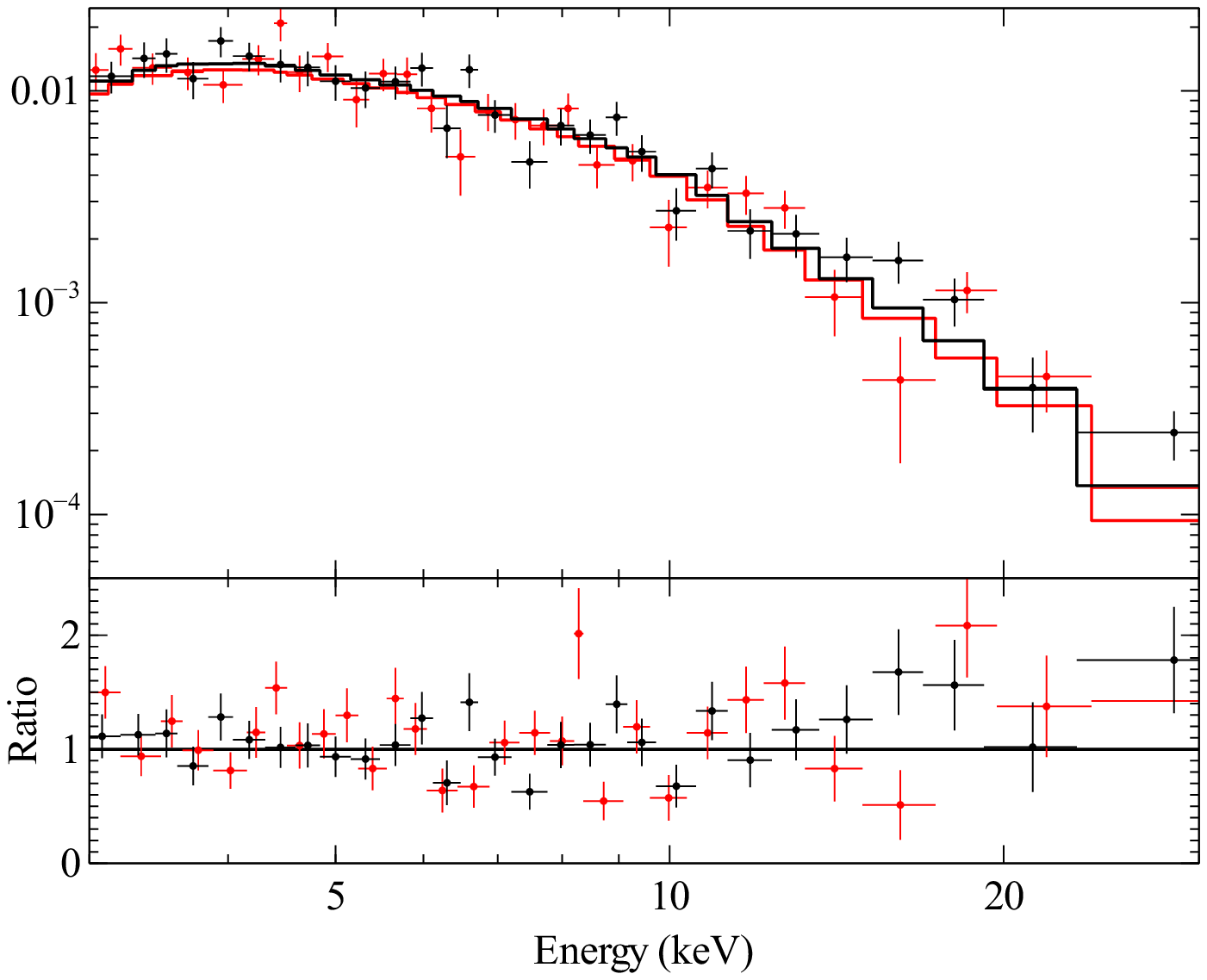}
\caption{Top: Difference spectrum calculated by using the lowest flux state as the background spectrum for the highest flux state spectrum, fit with a simple power law model. Black and red points correspond to the FPMA and FPMB spectra. Bottom: Residuals to this fit.}
\label{diffspec}
\end{figure}

In Fig.~\ref{diffspec} we show a difference spectrum, calculated by using the lowest flux state spectrum as the background for the highest. This spectrum is well fit ($\chi^2_\nu=265/247=1.07$) with a simple power law model, with $\Gamma=2.1\pm0.1$, consistent with the value from spectral fitting of the mean spectrum. To estimate the limit of the reflection contribution to the difference between the highest and lowest flux states, we again use the \textsc{relxill} model to fit these data. We fix the parameters to those found from the time-averaged spectrum, leaving only the normalisation and reflection fraction free. From this fit, we find that the reflection fraction, and hence the contribution of changes in the reflected emission to changes in the spectrum, is $R<0.16$. This suggests that the spectral variability in these spectra is dominated by changes in the geometry or position of the primary X-ray source, rather than in its net flux, as this would cause both the reflected and primary emission to increase equally.

%
%

We next apply the same three reflection models used in \S~\ref{section_avg}, allowing the normalisation of the blurred reflection and power law components to change in the \textsc{reflionx} model, the reflection fraction and normalization in the \textsc{relxill} model, and the source height and normalization in the \textsc{relxilllp} model. All three models give very good fits, with a slight preference towards the first two models (those without the emissivity profile fixed at the lamp-post solution).

\begin{table*}
\begin{tabular}{c c c c c c c c c  c}
\hline
Model & $N_\textrm{H}$ & $\log(\xi_\textrm{abs})$ & $\Gamma$ & $A_\textrm{Fe}$ &$\theta$&$a$&$\log(\xi_\textrm{ref})$&$q_\textrm{in}$ &$\chi^2_\nu/\textrm{d.o.f.}$\\
& $10^{22}$~cm$^{-2}$ & log(erg cm s$^{-1}$)& & solar&degrees&&&log(erg cm s$^{-1}$)&\\
\hline
\hline
1 & $10\pm4$ & $>3.78$ & $2.28\pm0.04$ & $2.2\pm0.2$ & $65\pm1$ & $0.98\pm0.01$ &$<1.7$&$>8$ &0.99/1078\\
2 & $10\pm3$ & $>3.88$ & $2.14^{+0.02}_{-0.04}$ & $2.6\pm0.2$& $65\pm1$ & $0.98\pm0.01$ &$2.0\pm0.2$ & $>9$ & 0.99/1078\\
3 & $16_{-7}^{+1}$ & $>3.72$ & $2.08_{-0.03}^{+0.02}$& $2.2\pm0.2$ & $42\pm2$ & $>0.994$ & $<1.16$ & - & 1.02/1078\\
\hline
\end{tabular}
\caption{Model parameters for fits to the flux-resolved \nustar\ spectra. All models include absorption by a single, highly ionised absorption zone, modelled with \textsc{xstar}. Model 1 is \textsc{reflionx}, model 2 is \textsc{relxill}, and model 3 is \textsc{relxilllp}.}
\label{resolvedfittable}
\end{table*}

\begin{figure}
\includegraphics[width=\linewidth]{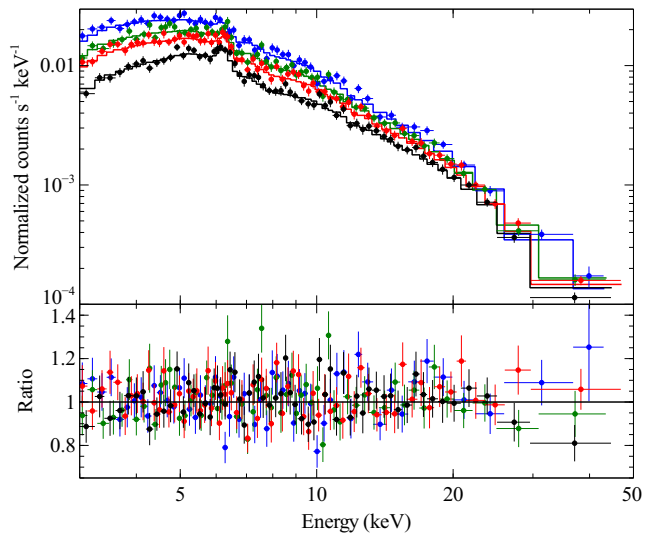}
\caption{Data and residuals from fitting the \textsc{relxill} model to the four flux state spectra. This fit is representative of the three reflection models. The FPMA and FPMB spectra are grouped and some binning is applied in \textsc{xspec} for clarity, but all spectra are fit separately and binned as described in the text.}
\label{fluxresolvedfit}
\end{figure}

We show the data and residuals to the best-fit \textsc{relxill} model in Fig.~\ref{fluxresolvedfit}. This is representative of all three reflection models gives an excellent fit to the data, and leaves no significant residuals in any of the spectra.

\begin{figure}
\includegraphics[width=\linewidth]{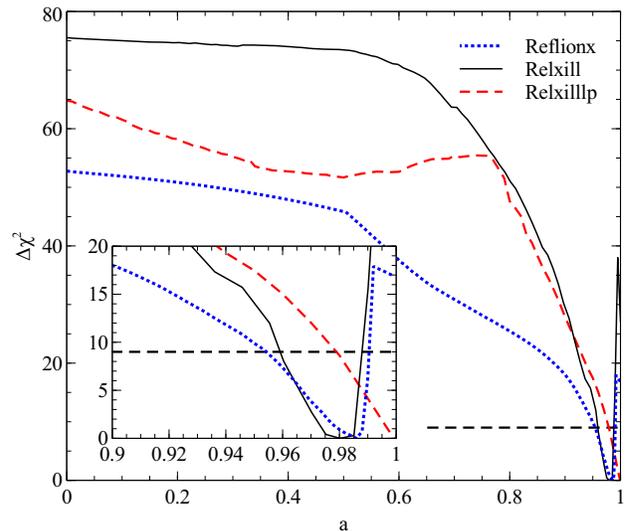}
\caption{$\Delta\chi^2$ contours for the three relativistic reflection models fit to the four flux-resolved spectra. The inset shows the $a=0.9$--1.0 region, and the dashed line shows the 3$\sigma$ confidence limit, for one parameter of interest.}
\label{resolvedchisquareds}
\end{figure}

As with the average spectrum, we restrict the reflection fraction to be less than the theoretical maximum for each spin value, again obtaining extremely strong spin constraints. The $\chi^2$ contours from the three blurred reflection models are shown in Fig.~\ref{resolvedchisquareds}, and as in Fig.~\ref{avgchisquareds} they show that high spin is preferred very strongly by all three models. The \textsc{reflionx} and standard \textsc{relxill} models also rule out maximal spin at high significance, unlike \textsc{relxilllp}. This is likely due to the additional constraints on the emissivity profile of the lamp-post model, which prevent it from finding the same optimal solution as the other two models. This may be pointing towards additional physics (such as ionisation gradients over this disk) or a difference in the coronal geometry from the pure lamp-post model. Alternatively, the other models may be over-fitting the data and the more constrained lamp-post model may be finding the more physical solution.

\begin{table}
\centering
\begin{tabular}{l c c c}
\hline
Flux & \textsc{reflionx} & \textsc{relxill} & \textsc{relxilllp}\\
& $R$ & $R$ & $h$ $(R_\textrm{G})$\\
\hline
\hline
Very low &$>7.9$&$>8.0$&$<2.3$\\
Low &$5.9_{-1.7}^{+3.9}$&$4.6_{-0.6}^{+0.8}$&$3.1_{-0.2}^{+0.5}$\\
High &$4.8_{-1.3}^{+3.5}$&$3.6_{-0.5}^{+0.6}$&$3.7_{-0.3}^{+1.0}$\\
Very High&$3.6_{-0.9}^{+2.6}$&$2.7_{-0.4}^{+0.5}$&$5.1_{-0.8}^{+2.2}$\\
\hline
\end{tabular}
\caption{Reflection fractions or source heights for the three relativistic reflection models for each of the four flux-resolved spectra. The fluxes in the \textsc{reflionx} model are calculated using \textsc{cflux} and include errors on the normalisation as well as the reflection fraction, leading to systematically larger errors.}
\label{Rtable}
\end{table}

In the case of the lowest flux spectrum we find that, without the introduction of the constraint on $R$ from the spin, the power law component is unconstrained. This means that the source spectrum can be described by a pure reflection model.

We find that the reflection fraction systematically decreases with flux, and hence the source height increases with flux in the lamp-post model. Table~\ref{Rtable} shows the reflection fractions and source heights we find, for each of the four spectra.


\subsection{Emissivity profile}

\begin{figure}
\centering
\includegraphics[width=\linewidth]{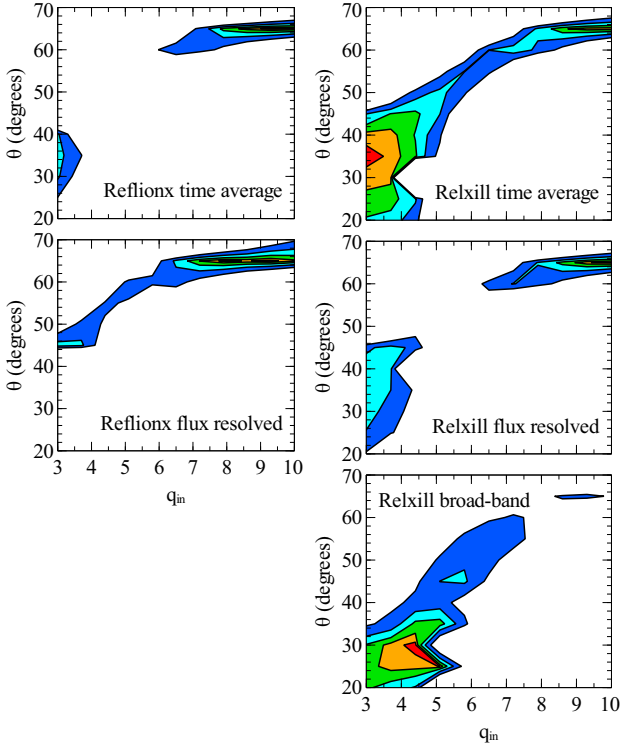}
\caption{1,2,3,4 and 5-$\sigma$ contours (red, orange, green, cyan and blue), showing the degeneracy between the inner emissivity index and the inclination, for the \textsc{reflionx} and \textsc{relxill} models, and the different datasets used. Generally, we find solutions with either low $\theta$ and low $q_\textrm{in}$, or high $\theta$ and high $q_\textrm{in}$. In the case of \textsc{relxill} fit to the time-average spectrum (top right) both solutions are acceptable.}
\label{qvsicontours}
\end{figure}

In several fits, we have found that there appear to be two acceptable values of the inclination and inner emissivity index. In Fig.~\ref{qvsicontours}, we show the $\chi^2$ contours for these two parameters, for all the fits presented so far. In general, we find that there are two acceptable solutions, one with high $q$ and high $\theta$, and one with low $q$ and low $\theta$. The \textsc{reflionx} models prefer both to be high, while the \textsc{relxill} models prefer the parameters to be high in the flux-resolved case, low in the broad-band average spectrum (although significantly higher than the classical limit of 3 \citet{Reynolds97}), and either solution is acceptable in the time-averaged case. This degeneracy likely arises from the limited spectral resolution, which makes it difficult to disentangle the blue edge of the broad line from the narrow line and absorption features. Since the blue edge of the line is the main driver of the inclination constraints, this parameter is highly sensitive to these effects, and our estimates should therefore be treated with a degree of caution. The differences between the two models could be due to the different treatment of angles - \textsc{reflionx} is angle-averaged, whereas \textsc{relxill} has an angle dependence in the reflection spectrum, or alternatively it could be due to the presence of the Fe~K$\beta$ line in the \textsc{relxill} model.

A more precise way of estimating the emissivity profile is given by \citet{Wilkins11}.  In brief, this method fits the iron line as a sum of line profiles from different annuli over the disk, then the relative normalisations of the lines from each annulus give the full emissivity profile. By restricting this to the 3-5~keV band (i.e., the red wing of the iron line) and fixing the power law parameters at the best fitting values, we can exclude the narrow line and absorption features from the fits, giving a more accurate constraint. This method can also allow for an independent constraint on the source height, separate from that obtained using the reflection fraction, as the precise shape of the line profile is strongly dependent on the position of the source.

\begin{figure}
\includegraphics[width=\linewidth]{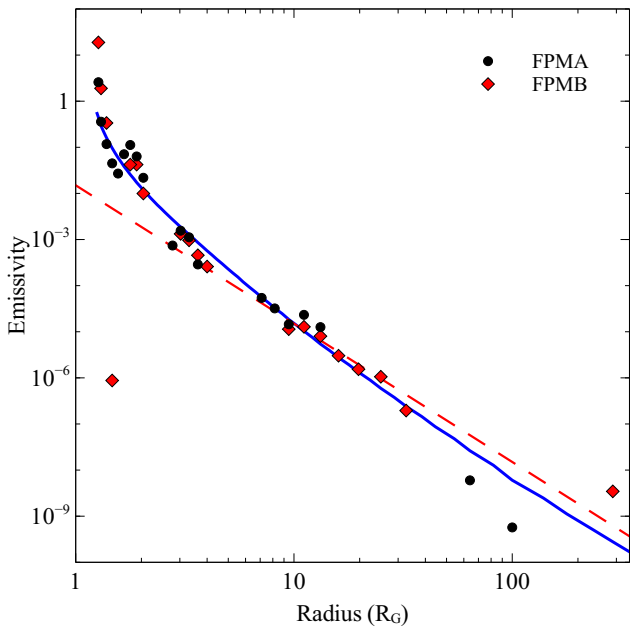}
\caption{Emissivity profile in arbitrary units calculated from the second, long observation of Mrk 335 for FPMA (black points) and FPMB (red points). The dashed red line shows the classical $q=3$ profile, and the solid blue line shows the theoretical emissivity profile for a point source at a height of 2~$R_\textrm{G}$ from the disk plane. These lines are not fit to the points, and are simply intended as a guide.}
\label{qprofile}
\end{figure}

Fig.~\ref{qprofile} shows the emissivity profile obtained using this method. At large radii the profile converges to the classical limit of $q$=3 (shown by the red dashed line), but at lower radii it is extremely steep. A profile of this shape is very similar to the theoretical profile for a point source very close to the disk plane \citep{Martocchia00}, and we show the prediction for a lamp-post source at 2~$R_\textrm{G}$ above the disk plane (blue solid line). The steepness of the line profile at small radii found using this method confirms the need for a large degree of light-bending to explain the observed spectrum. The missing values at intermediate  and large radii are due to the lack of a constraint on the emissivity at these points, as these regions of the disk to not contribute strongly to the red wing of the line profile \citep[for more detail, see][]{Wilkins11}.


While it seems likely that the emissivity profile is steep (no model significantly prefers a fit with $q<4$), as would be expected if the light-bending interpretation is correct, it difficult to be sure exactly how steep. The values of $\theta$ found for the high $q$ solutions may be unphysically high for a NLS1, and it may be that the true values of $\theta$ and $q$ are more moderate. It appears that the \suzaku\ XIS data, with their higher spectral resolution, favour such a solution (Gallo et al., in preparation).


\section{Discussion}
We have demonstrated that the broad-band spectrum of a low state in Mrk~335 can be well fit with relativistic reflection. We find a high spin, and very high reflection fractions, which suggest that the low state is caused by a collapse of the corona in towards the event horizon. 

These Mrk~335 observations appear very similar to the \emph{XMM-Newton} observation of 1H0707-495 in 2011, described by \citet{Fabian12}. \citeauthor{Fabian12} found that the spectrum could be described by blurred reflection alone, without a power law component (although one was slightly preferred by the fit, it was not constrained, as in the lowest flux spectrum of Mrk 335), and both Fe-K and Fe-L lines were clearly visible and strongly blurred. A similarly steep emissivity profile was obtained for 1H0707-495, as well as high spin. From the similarity of the source state and parameters, we conclude that these two objects are exhibiting the same phenomenon. The light bending solution is very similar to that found for MCG--6-30-15, where the variability is dominated by the   power law component, and the relativistic reflection is much more constant \citep{Parker13, Vaughan04, Fabian02,Miniutti03}. Similar light bending effects have also been seen in NGC 3783 \citep{Reis12}, IRAS 13224 \citep[Chiang et al., submitted]{Fabian13_iras, Kara13_iras}, and the XRB XTE~J1650-500 \citep{Rossi05, Reis13}.

We note that the $\sim150$~s Fe K lag observed by \citet{Kara13} was during a high flux state. This suggests that the lower edge of the corona must still have been close to the disk (within a few $R_\textrm{G}$), despite the increased flux, whereas the simple lamp-post model used here would suggest that the source height should be larger in a high flux state. It is possible that this conflict is due to the simplified geometry used, as the corona, while known to be compact, may not be well approximated by a point source. If we consider a vertically extended corona which contracts downwards towards the black hole, then the primary continuum emission will be dominated by the upper surface and the blurred reflection will be dominated by the lower surface. In this way the reflection spectrum would change very little, while the continuum would decrease sharply in flux. 

We have demonstrated the ability of the maximum reflection fraction method introduced in Dauser et al. (submitted) to constrain the spin parameter, obtaining consistent and precise measurements of the spin from different reflection models.
The spin estimate for the \textsc{relxilllp} model is extreme in both the average and flux-resolved spectra. We suggest that this may be due to the assumed emissivity profile, which does not take into account effects which can steepen the profile, such as changes in the ionisation parameter across the disk \citep{Svoboda12}. While this is a potential issue, this model still gives a very good fit, and is more physically motivated and constrained than the other models. More complex source geometries and determination of a more accurate ionisation profile are beyond the scope of this work, but are promising targets for the next generation of relativistic reflection models. 

Previous work on modelling the reflection spectrum of Mrk~335 has also found high spin values. \citet{Gallo13} found $a>0.7$ at 90 per cent confidence in an intermediate flux spectrum observed with \emph{XMM-Newton}, and \citet{WaltonSuzaku} found $a=0.8\pm0.1$ in a high flux state using \emph{Suzaku} XIS and PIN spectra. Our best fit value of $0.98\pm0.01$ confirms the high spin of this source, and greatly improves on the precision of this measurement. We note that there are further possible systematic effects which may affect the accuracy of this measurement, such as emission from within the innermost stable circular orbit \citep{Reynolds08}.

Spectra showing broad features around the iron line can generally be fit with either reflection or absorption models \citep[e.g][]{Miller13}. In this work we have focused on the reflection interpretation of these data, motivated by the presence of an Fe~K lag \citep{Kara13}, but we do not rule out the possibility that this spectrum could be produced by other models. We note that a detailed comparison of alternative models over the soft band using the \suzaku\ data (Gallo et al., in preparation) suggests that the reflection model is statistically preferred. Additionally, we stress that the light-bending model reproduces the spectral variability in a self-consistent manner and provides definite predictions, such as broadening of the line in a low state.

\section{Conclusions}
Detailed spectral analysis of an extremely low flux observation of Mrk 335 by \nustar{} has revealed spectra with exceptionally high fractions ($\gtrsim8$) of relativistically blurred emission, which we interpret as being due to the X-ray source collapsing down to within $\sim2$ $R_\textrm{G}$ of the black hole event horizon. Fitting with the \textsc{relxilllp} blurred reflection model, which self-consistently calculates the reflection fraction and blurring parameters for a point source above the disk suggests a change in the height (above the disk plane) from $<2.3R_\textrm{G}$ to $5.2_{-0.8}^{+2.2}R_\textrm{G}$ during the observations.

By fitting the spectra we find parameters indicative of reflection from the inner edge of the accretion disk around a rapidly spinning black hole, which is confirmed by analysis of how the reflection fraction changes with the source flux, and by determination of the emissivity profile, which is typical for a source within a few $R_\textrm{G}$ of the event horizon.
By excluding unphysical solutions we have calculated extremely precise measurements of the AGN spin, $a=0.98\pm0.1$, based on the requirement that a high spin is needed to produce the necessary light-bending for a reflection-dominated spectrum.

We conclude that the extreme spectral variability in Mrk 335 can be attributed to light-bending due to strong relativistic effects focusing the continuum radiation onto the disk at low source heights, and this observations yields a spectral image of the innermost region around a rapidly spinning black hole.

\section*{Acknowledgements}
This work is based on observations made by the \nustar{} mission, a project led by the California Institute of Astronomy, managed by the Jet Propulsion Laboratory, and funded by NASA. This research has made use of the \nustar{} Data Analysis Software (NuSTARDAS), jointly developed by the ASI Science Data Center (ASDC, Italy) and the California Institute of Technology (USA).
At Penn State \emph{Swift} is supported by NASA contract NAS5-00136.
MLP acknowledges financial support from the Science and Technology Facilities Council (STFC). 
ACF thanks the Royal Society for support.
AM and GM acknowledge financial support from Italian Space Agency under grant ASI/INAF I/037/12/0 - 011/13.
DRW is supported by a CITA National Fellowship.
The research leading to these results has received funding from the European Union Seventh Framework Programme (FP7/2007-2013) under grant agreement n.312789.
\bibliographystyle{mnras}
\bibliography{bibliography_mrk335}
\end{document}